\documentclass[10pt]{iopart}

\begin{document}

\title[Hemispherical anomalies]
{Constraints on cosmic hemispherical power anomalies from quasars}

\author{Christopher M Hirata}

\address{Caltech M/C 350-17, Pasadena, CA 91125, USA}

\ead{chirata@tapir.caltech.edu}

\begin{abstract}
Recent analyses of the cosmic microwave background (CMB) maps from the WMAP satellite have uncovered evidence for a hemispherical power
anomaly, i.e. a dipole modulation of the CMB power spectrum at large angular scales with an amplitude of $\pm14$ percent.  Erickcek \etal\ have put forward an
inflationary model to explain this anomaly.  Their scenario is a variation on the curvaton scenario in which the curvaton possesses a large-scale spatial
gradient that modulates the amplitude of CMB fluctuations.  We show that this scenario would also lead to a spatial gradient in the amplitude of perturbations
$\sigma_8$, and hence to a dipole asymmetry in any highly biased tracer
of the underlying density field.  Using the high-redshift quasars from the Sloan Digital Sky Survey, we find an upper limit on such a gradient of
$|\nabla\sigma_8|/\sigma_8<0.027r_{\rm lss}^{-1}$
(99\% posterior probability), where $r_{\rm lss}$ is the comoving distance to the last-scattering surface.  This rules out the simplest version
of the curvaton spatial gradient scenario.
\end{abstract}


\section{Introduction}

The simplest inflationary models for the origin of structure of the Universe has made a number of successful predictions, including the flatness of the Universe 
($\Omega_K=0$), and near-scale invariance ($n_s\approx 1$), adiabaticity, and Gaussianity of the primordial fluctuations.  Another prediction of these models is that 
the perturbations should be statistically homogeneous and isotropic.  It was therefore a surprise when, after the cosmic microwave background (CMB) maps from the 
Wilkinson Microwave Anisotropy Probe (WMAP) satellite were released \cite{2003ApJS..148....1B, 2007ApJS..170..288H, 2009ApJS..180..225H}, several groups reported 
evidence for large-scale anomalies in the data that were suggestive of inconsistencies with the paradigm of a statistically isotropic 
Gaussian random field, or with the theoretical power spectrum.  One widely discussed anomaly was an alignment of some of the low multipoles with a so-called Axis of 
Evil \cite{2003PhRvD..68l3523T, 2004PhRvD..70d3515C, 2005ApJ...618L..63H, 2005PhRvL..95g1301L}. Another was the CMB quadrupole $C_2$, which was unexpectedly low for 
the standard cosmological model \cite{2003PhRvD..68l3523T}.  A third anomaly was a difference in the observed CMB power spectrum between two hemispheres 
\cite{2004ApJ...605...14E, 2004MNRAS.354..641H, 2007ApJ...660L..81E, 2008arXiv0812.3795H, 2009ApJ...699..985H}.  A variety of theoretical explanations for these 
anomalies have been proposed \cite{2003JCAP...07..002C, 2004PhRvD..70h3003G, 2005ApJ...629L...1J, 2005PhRvD..72j3002G, 2006PhRvD..74d1301B, 2007PhRvD..75h3502A, 
2008MNRAS.389.1453B, 2008PhRvD..78l3520E}.

The hemispherical power anomaly is the main subject of this paper.  Its properties, as measured by Hoftuft \etal\ \cite{2009ApJ...699..985H}, can be summarized as 
follows: (1) The direction of greatest power is $\hat{\bi p}=(l,b)=(224^\circ,-22^\circ)\pm24^\circ$, near the South Ecliptic Pole.  (2) The amplitude of large-scale 
fluctuations ($\ell\le 64$) is $A=0.072\pm0.022$ ($3.3\sigma$), meaning that the amplitude of the CMB perturbations $\Delta_T$ is 1.07 times the mean value in the 
$\hat{\bi p}$ direction, and 0.93 times the mean value in the $-\hat{\bi p}$ direction.\footnote{The asymmetry in the $C_l$s is twice this, $\pm 14$ percent, since 
the power spectrum is the square of the amplitude.} (3) The asymmetry is present in at least three of the WMAP bandpasses (41, 61, and 94 GHz), and in the 
foreground-suppressed internal linear combination (ILC) map \cite{2009ApJS..180..265G}.  (4) Only limited information is available about the $\ell$-dependence of the 
asymmetry. There are indications that it continues at least to $\ell=80$ \cite{2009ApJ...699..985H} and perhaps beyond \cite{2008arXiv0812.3795H},
but other analyses have found no evidence for an amplitude asymmetry at the first acoustic peak \cite{2005PhRvD..71d3002D}.

The large-scale anomalies are a difficult subject, in part because there is no clean way to assess their true significance.  While the quoted 
significances sometimes exceed 99\%, some of the analyses are {\em a posteriori}. In order to make progress it is necessary to consider the theoretical 
explanations for the large-scale anomalies, and ask what new predictions these models yield \cite{2008PhRvD..77f3008D}.  For some of the proposed ideas that only affect the low 
multipoles, the CMB temperature field has reached its cosmic variance limit and further tests must await future CMB polarization data with lower noise 
and foreground residuals than WMAP.  On the other hand, one class of models intended to explain the hemispheric 
power asymmetry make a host of new predictions testable with near-future or even current data.  Erickcek \etal\ \cite{2008PhRvD..78l3520E} recently 
proposed a variation on the curvaton model \cite{1990PhRvD..42..313M, 1997PhRvD..56..535L, 2001PhLB..522..215M, 2002PhLB..524....5L}: an inflaton 
$\phi$, which dominates the energy density during inflation, and a curvaton $\sigma$, which is 
effectively massless during inflation but decays after inflation ends to produce an energy density $\propto\sigma^2$.\footnote{We have assumed here that the curvaton 
never contributes more than a small fraction of the energy density of the Universe; this is true in the Erickcek \etal\ model, but it is not true in all curvaton 
models.}  In the variation of Erickcek 
\etal\ \cite{2008PhRvD..78l3520E}, there was initially a large-scale gradient in $\sigma$; since the primordial density perturbations are proportional 
to $\delta(\sigma^2)\approx 2\bar\sigma\delta\sigma$ (assuming they are sourced by the curvaton; inflaton-induced density perturbations may also be 
significant), the amplitude of the density perturbations is statistically inhomogeneous, being proportional to the local value of $|\bar\sigma|$.  A 
uniform gradient in the background field $\bar\sigma$ then results in a dipole modulation of the CMB power spectrum in the direction of some unit 
vector $\hat{\bi p}\parallel\nabla|\bar\sigma|$, with an amplitude given by the fractional variation of $\bar\sigma$ across our horizon volume.

As noted by Erickcek \etal\ \cite{2008PhRvD..78l3520E}, the proposed model could be tested with future CMB data, especially with the upcoming {\it 
Planck} satellite, which will observe many more modes in the CMB than WMAP and should definitively confirm or reject the power spectrum asymmetry 
possibly seen by WMAP.  (The model could possibly even be constrained further by using the higher multipoles in WMAP.)
However such a model also makes predictions for large-scale structure: since the dipole asymmetry in the power spectrum varies 
slowly with wavenumber $k$, it generates a large-scale gradient in the amplitude of fluctuations $\sigma_8$.  Thus the abundance of rare massive haloes 
should contain a spatial gradient.  Thus the angular distribution of these massive haloes (or any luminous tracer thereof) should contain a dipole in 
the same direction $\hat{\bi p}$ singled out by the CMB.  Moreover, using the measured dipole anisotropy of the CMB power spectrum and our knowledge of 
the halo mass function, one can specifically predict the amplitude of these features as a function of redshift and halo mass.  Such features, if 
detected with the correct dependences and alignment of the CMB, would be a smoking gun for new physics, whereas nondetection of the large scale 
structure dipole would rule out the simplest version of the Erickcek \etal\ \cite{2008PhRvD..78l3520E} model.

The purpose of this paper is two-fold: first, to develop the formalism to test for a spatial gradient in $\sigma_8$; and second, to apply this 
formalism to the current large-scale structure data and measure $\nabla\sigma_8$.  Our primary motivation was the Erickcek \etal\ 
\cite{2008PhRvD..78l3520E} model, but we note that any early-Universe explanation for the hemispherical asymmetry is likely to produce $\nabla\sigma_8$ 
and hence our measurement should provide tight constraints.  [We note that a statistically anisotropic power spectrum $P({\bi k})$ cannot produce the 
dipole signature seen by Refs.~\cite{2004ApJ...605...14E, 2007ApJ...660L..81E, 2008arXiv0812.3795H, 2009ApJ...699..985H} because it is invariant under ${\bi 
k}\rightarrow -{\bi k}$.] For our 
measurement of $\nabla\sigma_8$, we use the high-redshift spectroscopic quasars from the Sloan Digital Sky Survey (SDSS) as described below.

\section{Theory}

\subsection{The model}

The Erickcek \etal\ \cite{2008PhRvD..78l3520E} model 
produces a local primordial fluctuation amplitude $\Delta_{\cal R}(k)$ that has a spatial gradient:
\begin{equation}
\Delta_{\cal R}(k,\bi r) = \Delta_{\cal R}(k,0) \left(1 + \bi p\cdot\frac{\bi r}{r_{\rm lss}}\right),
\end{equation}
where $r_{\rm lss}$ is the distance to the surface of last scattering.  This normalization was chosen so that in observations of the CMB the amplitude 
of variations is modulated by a factor $1+A\cos\theta$, where $\theta$ is the angle between the gradient direction $\hat\bi p$ and the line of sight 
$\hat\bi n$ and $A=|\bi p|$ (with minor modifications due to reionization; see below).  Constraints in this paper will be reported on the amplitude 
$A$ and direction $\hat\bi p$ of this modulation.

\subsection{Effect on CMB}

A large-scale gradient in the primordial fluctuation amplitude will produce a corresponding dipole in the CMB anisotropies.  On small scales, where the 
added power from the integrated Sachs-Wolfe (ISW) effect and reionization are negligible, the local CMB power spectrum in a particular region of the 
sky $\hat\bi n$ is
\begin{equation}
C_l^{\rm loc}(\hat\bi n) = C_le^{-2\Delta\tau(\bi n)}(1+\bi p\cdot\hat\bi n)^2,
\end{equation}
where $\Delta\tau(\bi n)$ is the change in optical depth in direction $\bi n$ relative to the reference model.  There is motivation for such a term to 
exist: if $\sigma_8$ is larger in some regions than others, the regions of higher $\sigma_8$ will form more early galaxies and we expect that they will 
reionize first.  In the case where $A$ is small, we may write:
\begin{equation}
C_l^{\rm loc}(\hat\bi n) \approx C_l\left( 1 + 2X\bi p\cdot\hat\bi n \right),
\end{equation}
where the factor
\begin{equation}
X = 1 - \frac{r_{\rm ri}}{r_{\rm lss}}\frac{\rmd\tau}{\rmd\ln\sigma_8}
\end{equation}
accounts for the increase of $\sigma_8$ and hence $\tau$ in direction $+\hat\bi p$ versus $-\hat\bi p$, and the factor $r_{\rm ri}/r_{\rm lss}$ acconts 
for the fact that the reionization surface is closer to us than the last scattering surface.  Reionization occurs during the matter-dominated era when the 
optical depth depends on the scale factor at reionization $a_{\rm ri}$ via $\tau\propto a_{\rm ri}^{-3/2}$.  Therefore
\begin{equation}
X = 1 + \frac{3\tau r_{\rm ri}}{2r_{\rm lss}} \frac{\rmd\ln a_{\rm ri}}{\rmd\ln\sigma_8}.
\end{equation}
From the WMAP 5-year parameter constraints we know that $\tau = 0.087\pm 0.017$ \cite{2009ApJS..180..306D} and $3\tau r_{\rm ri}/(2r_{\rm lss}) = 
0.092\pm 0.021$.  One does not know $\rmd\ln a_{\rm ri}/\rmd\ln\sigma_8$, as this depends sensitively on astrophysics.  If reionization is 
caused by massive haloes, so that the dependence of the reionization redshift on $\sigma_8$ is dominated by changes in the halo mass function, then 
we have $a_{\rm ri}\propto\sigma_8^{-1}$ because the halo mass function during the matter-dominated era is invariant under rescalings of $a$ and 
$\sigma_8$ that hold $\sigma_8a$ constant.  This case gives $\rmd\ln a_{\rm ri}/\rmd\ln\sigma_8=-1$, i.e. higher $\sigma_8$ gives earlier reionization.  
This number will be increased somewhat (i.e. made closer to zero) if one takes into account that at earlier times more photons are required to reionize 
the intergalactic medium because of recombinations.  It will also be increased if higher $\sigma_8$ implies a larger clumping factor, or more 
minihaloes that act as a sink for ionizing photons.  
While the astrophysics is unclear at present, we are fortunate that the multiplying factor of 
$0.09$ renders $X$ relatively insensitive to the details:
the $\rmd\ln a_{\rm ri}/\rmd\ln\sigma_8=-1$ case would imply $X\approx 0.91$, whereas the extreme case of assuming reionization at an epoch independent of $\sigma_8$
would give $X=1$.  For practical purposes $X=1$ with negligible correction (unless the anisotropy is measured at $\ge 10\sigma$) and we will not consider it further.

The power asymmetry analysis of Ref.~\cite{2007ApJ...660L..81E} using the 3-year WMAP data and a range of scales $2\le\ell\le40$ found an asymmetry of 
$A=0.114$ with the direction $\hat\bi p$ given by $(l,b)=(225^\circ,-27^\circ)$ and 99.1\% significance.  A more recent analysis by Hansen \etal\ 
\cite{2008arXiv0812.3795H} using the 5-year WMAP data and going out to $\ell_{\rm max}=600$ finds preliminary evidence for a direction of 
$(l,b)=(226^\circ\pm10^\circ,-17^\circ\pm10^\circ)$ at 99.6\% significance. Very recently,
Hoftuft \etal\ \cite{2009ApJ...699..985H} search the range of scales up to $\ell_{\rm max}=64$ and find a 3.3$\sigma$ dipole,
$A=0.072\pm0.022$ with $\hat{\bi p}=(l,b)=(224^\circ,-22^\circ)\pm24^\circ$ in the WMAP 5-year ILC map.
In the 61 GHz band, they carried the analysis up to $\ell_{\rm max}=80$ and found $A=0.070\pm 0.019$ (3.7$\sigma$) with $\hat\bi p=(235^\circ,-17^\circ)\pm22^\circ$.

An analysis by Donoghue \& Donoghue \cite{2005PhRvD..71d3002D} of the first acoustic peak finds a power spectrum asymmetry of $\eta=2A=0.02\pm0.02$ at $\ell\sim 220$ 
(assuming the direction $\hat\bi p$ given by Hansen \etal \cite{2004MNRAS.354..641H}), implying $A<0.03$ at $2\sigma$.  This is in tension with the low-$\ell$ 
results, and motivates further analysis to reduce the uncertainty on $A$.

One way to reduce the error bars on the CMB method would be to extend it to higher $\ell$ where there are more modes available to constrain the hemispherical 
asymmetry.  The number of CMB modes available is proportional to $\ell_{\rm max}^2$ and so we would expect the uncertainty in $A$ to decrease as $\propto \ell_{\rm 
max}^{-1}$; the Hoftuft \etal\ uncertainties are in rough agreement with this expectation.  Extrapolating this scaling, an analysis making use of the full WMAP data 
out to $\ell_{\rm max}\sim 500$ should be able to reach $\sigma(A)\sim 0.003$ if other sources of uncertainty (e.g. unmasked point sources) can be overcome.  We 
believe such an analysis would be very helpful, especially to constrain the scale dependence of the asymmetry.  However, in this paper we take an independent approach 
with similar sensitivity: we look for an asymmetry in large-scale structure.

\subsection{Effect on large scale structure}

The large-scale gradient in $\sigma_8$ also has an effect on the abundance of massive haloes and any objects that occupy them.  For objects at a 
distance $r$, the observed 2-dimensional number density varies across the sky as:
\begin{equation}
\frac{\delta N}{N}(\hat\bi n) = \frac r{r_{\rm lss}}\frac{\partial\ln N}{\partial\ln\sigma_8}\bi p\cdot\hat\bi n.
\end{equation}
Our major problem here is determining $\partial\ln N/\partial\ln\sigma_8$.  This problem has been considered in the context of non-Gaussianity searches 
\cite{2008PhRvD..77l3514D, 2008ApJ...677L..77M, 2008JCAP...08..031S}, which found
\begin{equation}
\frac{\partial\ln N}{\partial\ln\sigma_8} = \delta_c(b-1),
\label{eq:dcb}
\end{equation}
where $\delta_c=1.69$ and $b$ is the bias.  This formula assumed a universal mass function, as found in simulations, and assumed that the objects under 
study have a halo occupation distribution (HOD) that depends only on halo mass.  The dependence on $\sigma_8$ makes conceptual sense: very massive 
objects are highly biased $b(M)\gg 1$, and their abundance increases rapidly with $\sigma_8$; low-mass objects are antibiased $b(M)<1$ and are less 
abundant at high $\sigma_8$ because they will have merged; and the overall abundance of dark matter particles with $b=1$ does not depend on $\sigma_8$.
If only recent major mergers are occupied, then there is an additional suppression in Eq.~(\ref{eq:dcb}) because massive haloes formed earlier if 
$\sigma_8$ is increased.  In the extended Press-Schechter (ePS) formalism, $b-1$ should then be replaced by $b-1-\delta_c^{-1}=b-1.59$ 
\cite{2008JCAP...08..031S}; this exponent has been confirmed by re-scaling of simulations \cite{2005Natur.435..629S,
2008MNRAS.386..577F} which suggest that during the matter-dominated era the $\sigma_8$ dependence for recent major mergers ranges between $b-1.59$ and 
$b-1.65$ depending on the remnant mass and progenitor mass ratio \cite{2008JCAP...08..031S}.  In the case of the SDSS quasars it is not clear what is 
the correct way to populate haloes and so both limiting cases ($b-1$ and $b-1.6$) must be considered.  In practice the bias of the high-redshift 
quasars is so large ($\sim 10$) that the difference between these cases is negligible.

In practice the quasars occupy a range of redshifts and the observed dipole is a superposition: $\delta N/N=\bi d\cdot\hat\bi n$, where
\begin{equation}
\bi d = \delta_c\bi p \int \frac {r(z)}{r_{\rm lss}} [b(z)-1]P(z)\,dz,
\label{eq:a}
\end{equation}
with the replacement $b-1\rightarrow b-1.6$ in the recent major merger case.  Here $P(z)$ is the redshift distribution of the quasars, normalized to 
unity: $\int P(z)\,dz=1$.

\section{Application to SDSS quasars}

We now turn to empirical determination of $\nabla\sigma_8$ using the above formalism.  

\subsection{Choice of dataset}

The high-redshift spectroscopic quasars from SDSS were chosen for this study because:
\begin{enumerate}
\item They are very distant, which means that the same spatial gradient in $\sigma_8$ translates into a large absolute difference 
$\Delta\sigma_8$ across the sky.
\item They have wide angular coverage, which mimimizes large-scale structure noise and provides the leverage in all three Cartesian 
coordinates necessary to measure a dipole.
\item They are highly biased, so that a small change in $\sigma_8$ is amplified into a much larger change in the 
number density of quasars.
\item They have very good rejection of Galactic objects (e.g. halo stars), which could otherwise produce spurious large 
angular scale modulation and potentially correlate with foreground residuals in the CMB maps.
\item The relative photometric calibration of SDSS is understood at the percent level (although this degrades to a few percent in the $u$ band).
\item The number density of $\sim 1$ deg$^{-2}$, while not impressive, is acceptable given the above advantages.  Given the choice between high number 
density of nearby objects with a weak dependence on $\sigma_8$, or a low number density of distant objects that depends strongly on $\sigma_8$, our 
preference is for the latter since Nature does much of the work to suppress systematics.
\end{enumerate}

There are other data sets that could have been used instead to trace large-scale structure at moderate to high redshifts.  Possibilities include:
\begin{enumerate}
\item {\em Luminous red galaxies}: These are abundant but current large-scale maps go out to only $z<0.7$ \cite{2007MNRAS.378..852P}, or 18\% of the 
distance to the last-scattering surface, and the typical bias is small, $b\sim 1.8$.
\item {\em Photometrically selected quasars}: These have a higher number density ($\sim$50 deg$^{-2}$), but at present the UV-based selection 
algorithms restrict this sample to $z<2.5$ \cite{2004ApJS..155..257R}.  The bias is also much lower at these redshifts, $b\sim$2.3--2.8 
\cite{2008PhRvD..78d3519H}.
\item {\em Hard X-ray background}: The 2--8 keV HEAO map has excellent sky coverage and perhaps the best-constrained dipole 
\cite{2002ApJ...580..672B}, but the 
redshift distribution is poorly known and the estimated bias is small and may be consistent with $b=1$ \cite{2004ApJ...612..647B}, so it is not known how 
sensitive this is to $\nabla\sigma_8$.
\item {\em Radio sources}: The NRAO VLA Sky Survey \cite{1998AJ....115.1693C} sample has been used to probe the moderate redshift range $z\sim 1$, but 
its redshift distribution is still under debate \cite{2004ApJ...608...10N, 2006PhRvD..74d3524P, 2008PhRvD..78d3519H}.  More importantly the maps 
exhibit declination-dependent striping which is almost certainly a systematic artifact and precludes 
determination of the dipole \cite{2002PhRvL..88b1302B}.
\item {\em Lyman-$\alpha$ forest}: A different approach would be to measure a fluctuation amplitude from the Lyman-$\alpha$ forest at the same redshift 
but with sightlines in different parts of the sky.  The differential nature of the measurement would suppress many of the astrophysical systematics 
in e.g. $P(k)$ measurement from the Lyman-$\alpha$ forest.  This approach was not taken here because the analysis would be much more 
complicated, but given the large number of observed modes in the Lyman-$\alpha$ forest it should be considered in the future.
\end{enumerate}

\subsection{Data description}

The high-redshift quasars used here are obtained from the SDSS.  The SDSS drift-scans the sky in five bands ($ugriz$) \cite{1996AJ....111.1748F} under 
photometric conditions \cite{2000AJ....120.1579Y, 2001AJ....122.2129H} using a 2.5-meter optical telescope \cite{2006AJ....131.2332G} with 3 degree 
field of view camera \cite{1998AJ....116.3040G} located in New Mexico, USA \cite{2000AJ....120.1579Y}. The photometric and astrometric calibration of 
the SDSS and the quality assessment pipeline are described by Refs.~\cite{2002AJ....123.2121S, 2006AN....327..821T, 2008ApJ...674.1217P, 
2003AJ....125.1559P,2004AN....325..583I}.  Bright galaxies \cite{2002AJ....124.1810S}, luminous red galaxies (LRGs) \cite{2001AJ....122.2267E}, and 
quasar candidates \cite{2002AJ....123.2945R} are selected from the SDSS imaging data for spectroscopic follow-up \cite{2003AJ....125.2276B}.

The spectroscopic quasar sample used here was based on colour selection criteria described in Richards \etal\ 
\cite{2002AJ....123.2945R}.  The algorithm selects quasar candidates down to a limiting magnitude of $i=19.1$ or $20.2$ (for high-$z$ quasars);
a discussion of its completeness and efficiency can be found in Richards \etal\
\cite{2002AJ....123.2945R}.
A statistical sample of confirmed 
quasars was generated by Shen \etal \cite{2007AJ....133.2222S}, who removed from the sample regions with early versions of the target selection 
algorithm.  Shen \etal imposed a further cut on the quality of the imaging data used for target selection, dividing it into ``good'' and ``bad'' 
fields.  As a test for systematics, their quasar clustering analysis was repeated with both the ``good'' fields only, and with ``all'' fields.

The Shen \etal \cite{2007AJ....133.2222S} sample has a minimum redshift $z\ge 2.9$, due to the need in target selection to
avoid Galactic late A/early F stars which have similar broadband
colours to quasars at $z\sim 2.7$, and most objects are at $z<4.5$ due to the declining luminosity function and increasing luminosity distance at high 
redshift.  There are 4426 quasars in the ``all'' sample (4041 deg$^2$) and 3846 in ``good'' (3506 deg$^2$).  The quasars undersample the density field 
at all scales, so that their power spectrum is dominated by Poisson noise.

Shen \etal specify a mask, which we have imported into {\sc Mangle} \cite{1993ApJ...417...19H, 2004MNRAS.349..115H, 2008MNRAS.387.1391S}.  {\sc Mangle} is a suite of 
computer programmes widely used in the large-scale structure community to manipulate complex masks, which it typically represents as spherical polygons (i.e. regions 
bounded by circular arcs).  {\sc Mangle} can carry out simple tasks on these masks such as unions and spherical harmonic transform computation; most importantly for 
us, it can generate catalogues of random points from within a given mask.

For this analysis, we have further split the quasars into a $2.9\le z<3.5$ sample, and a $3.5\le z<4.5$ sample.

\subsection{Dipole anisotropy estimator}

In general the quasar density in a given direction can be written as:
\begin{equation}
\frac{\delta N}{\bar N}(\hat\bi n) = \bi d\cdot\hat\bi n + \sum_i k_it_i(\hat\bi n) + C.
\end{equation}
Here $\bar N$ is the mean number of quasars per steradian, $\delta N$ is its fluctuation,
$\bi d$ is the dipole (which we wish to measure), $t_i(\hat\bi n)$ are possible systematics templates in the quasar maps (e.g. the extinction 
map), $k_i$ are the sensitivities of the quasar densities to these systematics, and $C$ is a mean offset over which we must marginalize since the true 
mean number density of quasars is not known.  This equation can be simply written as:
\begin{equation}
\frac{\delta N}{N}(\hat\bi n) = \bi x\cdot\bi T(\hat\bi n),
\end{equation}
where we have defined the vectors $\bi x = (\bi d,k_i,C)$ and $\bi T(\hat\bi n) = (\hat\bi n, t_i(\hat\bi n),1)$.  If Poisson noise dominates the 
uncertainty in 
$\delta N/\bar N$, so that the inverse variance per steradian is $\bar N$, then the best linear unbiased estimator for $\bi x$ is obtained via:
\begin{equation}
\hat\bi x = {\mathbf F}^{-1}{\bi g},
\label{eq:fg}
\end{equation}
where
\begin{equation}
g_i = \int T_i(\hat\bi n)\,\delta N(\hat\bi n) d^2\hat\bi n
\end{equation}
and the Fisher matrix is
\begin{equation}
F_{ij} = \bar N \int T_i(\hat\bi n) T_j(\hat\bi n)\,d^2\hat\bi n.
\end{equation}
The covariance matrix of this estimator is ${\rm Cov}(\hat\bi x)={\mathbf F}^{-1}$.

The above equations are integrals over the survey area, but quasars are discrete objects and so it is easiest to replace the above results with 
summations over the quasars and random catalogues.  We can do this by writing:
\begin{equation}
g_i = \sum_{\rm D} T_i(\hat\bi n_{\rm D}) - \frac{N_{\rm D}}{N_{\rm R}} \sum_{\rm R} T_i(\hat\bi n_{\rm R}),
\end{equation}
where $\hat\bi n_{\rm D}$ are the positions of the quasars in the real data, $\hat\bi n_{\rm R}$ are the positions of randomly generated points in 
the survey mask, and $N_{\rm D}$ and $N_{\rm R}$ are the number of quasars (data) and random points respectively.  The Fisher matrix is:
\begin{equation}
F_{ij} = \frac{N_{\rm D}}{N_{\rm R}} \sum_{\rm R} T_i(\hat\bi n_{\rm R}) T_j(\hat\bi n_{\rm R}).
\end{equation}
We have generated random catalogues containing $N_{\rm R} = 10^6$ points using {\sc Mangle}'s {\sc ransack} task \cite{2008MNRAS.387.1391S}, which is more than 
sufficient as their residual noise is negligible ($<1$\%) compared to the actual data.

\subsection{Measured dipole anisotropy}

We have applied the estimator Eq.~(\ref{eq:fg}) to the high-$z$ SDSS quasar data.  Four versions of the fit were done, which differed by using the 
`all'' or ``good'' field mask, and by either including or not including the dust template.  The dust template used is the predicted reddening map of 
$E(B-V)$ by Schlegel \etal \cite{1998ApJ...500..525S} as a systematics template, which is appropriate if the extinction correction is not perfect.  
Errors correlated with the map could arise due to errors in the assumed extinction law $A_\lambda/E(B-V)$, miscalibration of the $E(B-V)$ scale, or 
unusual quasar colours affecting the extinction correction (e.g. if a bright emission line lies at the red end of the bandpass then the band-averaged 
correction may overcorrect for the extinction).  The template amplitude $k_{E(B-V)}$ could plausibly be positive or negative depending on the amount of over- or 
undercorrection in different bands.  Because most of the data lie in the Northern Galactic cap, there is a strong degeneracy between the component of 
the dipole perpendicular to the Galactic plane and the extinction correction, Corr$(k_{E(B-V)},d_z)=0.65$: either increasing $d_z$ or decreasing $k_{E(B-V)}$ would 
have 
the effect of enhancing the number density of objects near the NGP.  This degeneracy could be broken in the future with more data in the southern 
patch.  For the ``all'' mask, we find template amplitudes of $k_{E(B-V)}=-1.6\pm1.2$ (low-$z$) and $k_{E(B-V)}=-3.5\pm1.6$ (high-$z$); these become $-1.0\pm1.3$ and 
$-3.6\pm1.7$ for the ``good'' mask.  We regard the fits with the dust template as our primary result because correlation with dust is
marginally detected at $>2\sigma$ in one of the redshift slices.

The resulting dipole moments in the Galactic coordinate system are shown in Table~\ref{tab:d-obs}.  To be explicit, we have placed the $x$ direction 
toward the Galactic centre, the $y$ direction in the plane at $l=90^\circ$, and the $z$ direction toward the North Galactic Pole (NGP).  Using the 
``all'' mask and no dust template, one finds an intriguing $2.9\sigma$ hint of $d_z>0$ in the low-$z$ quasars.  However the fact that the preferred 
direction is very near the NGP ($b=80^\circ$) is suspicious, and indeed this signal goes away if one restricts to ``good'' fields (significance drops 
to 1.8$\sigma$), uses the dust template (1.4$\sigma$), or both (0.8$\sigma$).  

In all cases the uncertainty in the dipole determination is at least several percent, so no correction for our own peculiar velocity $v/c=0.0012$ 
\cite{2007ApJS..170..288H} is necessary.

\begin{table}
\caption{\label{tab:d-obs}The observed dipole anisotropies, in Galactic Cartesian coordinates, for the various quasar samples and for the dust 
template either included (Yes) or not (No).}
\lineup
\begin{indented}
\item[]\begin{tabular}{@{}lllll}
\br
Sample & Dust? & $10^2d_x$ & $10^2d_y$ & $10^2d_z$ \\
\mr
2.9--3.5, all & No & $-\01.3\pm5.1$ & $\m\02.6\pm6.5$ & $\m16.3\pm\05.6$ \\
3.5--4.5, all & No & $-\00.5\pm6.7$ & $-\04.2\pm8.4$ & $-\01.9\pm\07.3$ \\
2.9--3.5, all & Yes & $\m\01.1\pm5.4$ & $-\00.7\pm6.9$ & $\m10.0\pm\07.4$ \\
3.5--4.5, all & Yes & $\m\04.8\pm7.1$ & $-11.7\pm9.0$ & $-15.9\pm\09.6$ \\
2.9--3.5, good & No & $-\02.3\pm5.5$ & $\m\02.8\pm6.9$ & $\m10.8\pm\06.0$ \\
3.5--4.5, good & No & $\m\03.2\pm7.1$ & $-\02.7\pm9.0$ & $-\07.4\pm\06.8$ \\
2.9--3.5, good & Yes & $\m\03.9\pm5.8$ & $\m\00.6\pm7.5$ & $\m\06.6\pm\07.9$ \\
3.5--4.5, good & Yes & $\m\08.7\pm7.6$ & $-10.5\pm9.6$ & $-21.9\pm10.2$ \\
\br
\end{tabular}
\end{indented}
\end{table}

In addition to the Galactic dust, a second concern would be large-angle calibration errors, which can involve either ``grey'' errors (in which all five bands move together) or colour calibration errors. 
We may do a simple assessment of the grey calibration error by comparing the target selection photometry to the ubercalibrated photomety \cite{2008ApJ...674.1217P}
obtained from the Shen \etal\ \cite{2007AJ....133.2222S} catalogue, in $i$ band as used for the flux cut.  
The ubercalibrated photometry is estimated to be accurate to 1\%.  We compare the magnitude difference $\Delta i = i_{\rm target}-i_{\rm ubercal}$ for each quasar, and do an unweighted least-squares fit
\begin{equation}
\Delta i = \Delta i_0 + \bi m\cdot\hat\bi n,
\end{equation}
where $\bi m$ is the calibration dipole and $\bi n$ is the position of that quasar.
We remove two (possibly variable) objects from the fit that were undetected in the ubercalibrated photometry.
The resulting dipole is $10^3\bi m = (0\pm3, -10\pm3, -3\pm3)$.
We have also tried clipping the distribution of $\Delta i$ at the 1st and 99th percentiles (i.e.
quasars below the 1st percentile were moved to the 1st percentile value before fitting).  The resulting dipole is $10^3\bi m = (0.6\pm 0.8, -3.5\pm0.9, 0.5\pm0.8)$.
The implied dipole due to changes in the $i$ band calibration would be 1.8 times this, since
the slope of the cumulative quasar luminosity function
for this redshift range and the $i<20.2$ flux cut is $\sim 1.8$ (see e.g. Fig. 13 of Ref.~\cite{2006AJ....131.2766R}).
Even the 1\% dipole obtained from the first fit would lead to a calibration-induced dipole of 1.8\% in the density of quasars, which is much
less than our 5--10\% statistical errors.  (But do note that both calibration dipoles are inconsistent with zero, giving $\chi^2$ for 3 degrees of freedom of 12 and 19, respectively.)

Colour calibration is potentially nastier since quasars are selected by cuts that run through
the quasar locus, and hence could result in completeness variations if there is a relative colour calibration error.  Our best
test of the colour-dependent calibration will be the agreement (or lack thereof) of the dipoles inferred from different redshift
ranges where the colour selection is very different.  However, we can also test for this by
computing the dipole of the reddening-corrected colours, i.e. doing an unweighted least-squares fit
\begin{equation}
u-g = (u-g)_0 + \bi m\cdot\hat\bi n,
\end{equation}
where $(u-g)_0$ and the vector $\bi m$ represent 4 free parameters.  We use all of the quasars in the $2.9\le z<4.5$ range for this test.
Since some quasars are undetected in $u$, we clip the distribution at the 5th and 95th percentiles in $u-g$ before performing the fit (i.e.
quasars below the 5th percentile were moved to the 5th percentile value before fitting).  One can then test for whether $\bi m\neq 0$, which
would possibly indicate a dipole variation in colour calibration across the sky.  We report the $\chi^2$ value, $\bi m\cdot{\rm Cov}_{(\bi m)}\bi m$,
and the probability to exceed $P(>\chi^2)$.  Similar tests can be done for the other colours, and for
the extinction-corrected apparent magnitudes of the quasars.  The magnitudes and colours used are those used for the target selection (TARGET PSF
photometry), which were obtained via matching to the SDSS DR5 quasar catalogue \cite{2007AJ....134..102S}.
The results are shown in Table~\ref{tab:dip}.  As one can see, the $\chi^2$ values
for these fits are all acceptable, with one possible exception ($r-i$, $p = 0.028$).  Given that we searched 10 directions in
colour space, the presence of this one anomalously high $\chi^2$ value is not a concern.

\begin{table}
\caption{\label{tab:dip}The $\chi^2$ values and probabilities to exceed for searches for dipole gradients in the quasar colour.}
\lineup
\begin{indented}
\item[]\begin{tabular}{@{}lll}
\br
Quantity & $\chi^2(\bi m)$ & $P(>\chi^2)$ \\
\mr
$u-g$ & 3.05 & 0.38 \\
$g-r$ & 4.61 & 0.20 \\
$r-i$ & 9.10 & 0.028 \\
$i-z$ & 2.84 & 0.42 \\
$u-r$ & 2.22 & 0.53 \\
$g-i$ & 4.53 & 0.21 \\
$r-z$ & 6.29 & 0.10 \\
$u-i$ & 1.58 & 0.66 \\
$g-z$ & 4.92 & 0.18 \\
$u-z$ & 1.36 & 0.72 \\
\br
\end{tabular}
\end{indented}
\end{table}

\subsection{Quasar bias and final results}

In order to convert the measured dipole anisotropy into a constraint on the perturbation amplitude asymmetry $A$ using Eq.~(\ref{eq:a}), we need to know the bias of the 
quasars.  If one assumes a cosmological model, one may obtain this by fitting to the correlation function.  Here we use the WMAP5+BAO cosmology, based 
on combining the WMAP 5-year power spectrum \cite{2009ApJS..180..296N} with the SDSS+2dF baryon acoustic oscillation (BAO) constraint 
\cite{2007MNRAS.381.1053P}; this combination yields $\sigma_8=0.807\pm 0.027$ \cite{2009ApJS..180..306D}.

If one has a measured correlation length $r_0$ and an assumed cosmology, the bias can be obtained by the formula $b=[\xi_m(r_0)]^{-1/2}$, where 
$\xi_m(r)$ is the matter correlation function.  We have used the linear correlation function using the Eisenstein \& Hu transfer function 
\cite{1998ApJ...496..605E} since in the regime of interest the nonlinear corrections as estimated by the Smith \etal \cite{2003MNRAS.341.1311S} 
prescription are $<1$\%, i.e. negligible.  The correlation lengths that Shen \etal \cite{2007AJ....133.2222S} measured from the ``good'' fields are $17.91\pm 1.50$ 
$h^{-1}\,$Mpc 
($2.9\le z<3.5$) and $25.22\pm 2.50$ $h^{-1}\,$Mpc ($z\ge 3.5$).  These correlation lengths imply $b=8.9\pm 0.7$ ($z=3.2$) and $b=15.3\pm 1.7$ 
($z=4.0$), where the uncertainty is simply propagated from $r_0$.  There is some additional uncertainty in these numbers from nonlinear biasing, from 
the cosmological parameters, and because the correlation length does not appear to be constant with redshift.  We note that the one-halo term in the 
correlation function cannot contaminate these measurements because Shen \etal only fit projected separations $r_p>4h^{-1}\,$Mpc.  Integrating over the 
redshift bin assuming a constant correlation length $r_0$, we turn Eq.~(\ref{eq:a}) into:
\begin{equation}
\bi d = \left\{\begin{array}{rll}
(5.8\pm0.6)\bi p & & {\rm low-}z{\rm~slice} \\
(11.5\pm1.4)\bi p & & {\rm high-}z{\rm~slice}
\end{array}\right..
\label{eq:response}
\end{equation}

Using the central value of the conversion from the observed quasar dipole into $\bi p$, and the conservative (good + dust marginalization) dipole 
estimates, we obtain $10^2\bi p = (0.7\pm 1.0, 0.1\pm 1.3, 1.1\pm 1.4)$ for the low-$z$ slice and $(0.8\pm0.7,-0.9\pm0.8,-1.9\pm0.9)$ for the high-$z$ 
slice.  Combining the two results with the usual weighting by their inverse-covariance matrix gives
\begin{equation}
10^2\bi p = (0.73, -0.61, -1.01),
\label{eq:central}
\end{equation}
with covariance:
\begin{equation}
10^4{\rm Cov}_{(\bi p)} = \left(\begin{array}{rrr}
0.306 & -0.225 & -0.122 \\
-0.225 & 0.497 & 0.227 \\
-0.122 & 0.227 & 0.560
\end{array}\right).
\end{equation}
In the optimal determination of the central value (Eq.~\ref{eq:central}) the low-redshift slice receives 30\% of the weight and the high-redshift 
slice receives 70\%, i.e.
\begin{equation}
\bi p = w_{\rm lo-z}\bi p_{\rm lo-z} + w_{\rm hi-z}\bi p_{\rm hi-z}
\label{eq:weights}
\end{equation}
with $w_{\rm lo-z}=0.30$ and $w_{\rm hi-z}=0.70$.

The difference between the inferred dipoles from the two samples is $\bi p_{\rm lo-z}-\bi p_{\rm hi-z}=(-0.1,1.0,3.1)\times 10^{-2}$; based on the
covariance matrix for the difference, the $\chi^2$ is 3.03 (3 degrees of freedom; $P=0.39$) so we conclude the dipoles measured from the two redshift 
slices are consistent.  There is no detection of any global dipole ($\chi^2=2.80$; $P=0.42$).

If one fixes $\hat\bi p$ to be in the specific direction $(l,b)=(225^\circ,-27^\circ)$ identified by Eriksen \etal \cite{2007ApJ...660L..81E}, then 
the magnitude $A$ of $\bi p$ is found to be $10^2A=-0.18\pm0.44$, i.e. $-0.0105<A<0.0070$ at 95\% confidence ($-0.0132<A<0.0097$ at 99\%).  A 
directon-independent upper limit can be set by a 
Bayesian analysis analogous to 
Eriksen \etal: by placing a uniform prior on the magnitude $A$ and direction $\hat\bi p$, we can construct a marginalized likelihood function for $A$:
\begin{equation}
{\cal L}(A) \propto \int \exp\left[ -\frac12(\bi p-\bi p_{\rm best})\cdot{\rm Cov}^{-1}(\bi p-\bi p_{\rm best}) \right] d^2\hat\bi p,
\label{eq:lp}
\end{equation}
where $\bi p=A\hat\bi p$.
We find that 95\% of the Bayesian posterior distribution is at $A<0.019$ and 99\% at $A<0.026$.\footnote{Since $A$ is the magnitude of $\bi p$, in our 
direction-independent constraint it is required to be non-negative.  In the fixed-direction constraint where $\hat\bi p$ is fixed to the Eriksen \etal\ direction, we 
allow for the possibility of negative $A$, i.e. a dipole opposite to that seen in the CMB.}

In reality the above computations should take into accounte the uncertainty in the estimator response, Eq.~(\ref{eq:response}), which derives from the 
uncertainty in the quasar bias.  In particular it is possible to circumvent the above limits on $\bi p$ if the quasar bias is smaller than central 
values.  For our chosen 
weights, Eq.~(\ref{eq:weights}), the response of our estimator is:
\begin{equation}
\bi p_{\rm estimated} = (1.00\pm 0.09)\bi p_{\rm actual},
\end{equation}
i.e. there is a 9\% ($1\sigma$) calibration uncertainty.  We can include this uncertainty in a Bayesian analysis by incorporating a calibration factor 
$\chi$ with a prior at $1.00\pm 0.09$.  For the fixed-direction case (i.e. where $\hat\bi p$ is required to point in the Eriksen \etal direction), one 
may define a margimalized likelihood function for $A$:
\begin{equation}
{\cal L}_{\rm marg}(A) \propto \int \rme^{-(\chi A+0.0018)^2/2(0.0044)^2} \rme^{-(\chi-1)^2/2(0.09)^2} \rmd\chi.
\end{equation}
This gives $-0.0108<A<0.0071$ (95\%) or $-0.0138<A<0.0100$ (99\%).
For the direction-independent case,
\begin{equation}
{\cal L}_{\rm marg}(A) = \int {\cal L}_{\rm un-marg}(\chi A) \rm e^{-(\chi-1)^2/2(0.09)^2} d\chi,
\end{equation}
where the un-marginalized function ${\cal L}_{\rm un-marg}(A)$ is given by Eq.~(\ref{eq:lp}).  Here we find that 95\% of the Bayesian posterior 
distribution is at $A<0.020$ and 99\% is at $A<0.027$.  These are negligible changes from the un-marginalized case.

We have also repeated the above analysis for the ``all'' fields.  In this case, the correlation length measured by Shen \etal \cite{2007AJ....133.2222S} is shorter, 
so the 
biases are less: $b=8.0\pm0.9$ at $z=3.2$ and $b=13.5\pm1.6$ at $z=4.0$.
The dipole response is now
\begin{equation}
\bi d = \left\{\begin{array}{rll}
(5.1\pm0.7)\bi p & & {\rm low-}z{\rm~slice} \\
(10.0\pm1.3)\bi p & & {\rm high-}z{\rm~slice}
\end{array}\right..
\label{eq:response-all}
\end{equation}
The best-fit dipole and its covariance are:
\begin{equation}
10^2\bi p = (0.40, -0.87, -0.52)
\end{equation}
and
\begin{equation}
10^4{\rm Cov}_{(\bi p)} = \left(\begin{array}{rrr}
0.351 & -0.259 & -0.141 \\
-0.259 & 0.571 & 0.261 \\
-0.141 & 0.261 & 0.643
\end{array}\right).
\end{equation}
The weights in this case are $w_{\rm lo-z}=0.30$ and $w_{\rm hi-z}=0.70$.  Again the low-$z$ and high-$z$ slices are consistent ($\chi^2=3.11$, 
$P=0.37$).  There is also no detection of $\bi p$: the $\chi^2$ for zero dipole is 1.34 ($P=0.72$).
The calibration uncertainty is now 10\%.  After marginalizing over this, the fixed-direction constraint on $A$ is $-0.0073<A<0.0120$ (95\%) or 
$-0.0105<A<0.0153$ (99\%), and the direction-independent constraint is $A<0.017$ (95\%) or $A<0.025$ (99\%).

\subsection{Scale of constraint}

Up until now we have assumed that the dipole anisotropy $\hat\bi p$ is scale-invariant.  However inflationary models usually predict slight deviations 
from scale invariance because as one approaches the end of inflation each e-fold is slightly different from the previous one.  In order to constrain 
such models, we need to know the effective scale at which the quasars constrain $\hat\bi p$.  Since the fluctuation amplitude is typically a 
smoothly varying function of the number of e-folds or equivalently of $\ln k$, only a rough estimate of $k_{\rm eff}$ is required.

We now obtain an estimate of the scale probed by the quasars, i.e. at what scale $k_{\rm eff}$ we have actually constrained the large-scale 
gradient of the fluctuation amplitude.
The abundance of very massive haloes (i.e. well above the nonlinear mass) is roughly determined by the 
top-hat variance $\sigma(R)$, where $M=4\pi\bar\rho_{m0} R^3/3$ is the halo mass, $\bar\rho_{m0}$ is the present-day matter density, and $R$ is the 
comoving radius of the top-hat filter.  This is:
\begin{equation}
\sigma^2(R) = \int \Delta^2(k) W^2(kR)\,\rmd\ln k,
\end{equation}
where the window function is $W(x) = 3j_1(x)/x$.  Differentiating gives
\begin{equation}
\delta\ln\sigma(R) = \frac{\int \Delta^2(k)W^2(kR) \delta\ln\Delta(k)\,\rmd\ln k}
  {\int \Delta^2(k) W^2(kR)\,\rmd\ln k}.
\end{equation}
This can be approximated as $\delta\ln\sigma(R) = \delta\ln\Delta(k_{\rm eff})$, where:
\begin{equation}
\ln k_{\rm eff} = \frac{\int \Delta^2(k)W^2(kR) \ln k\,\rmd\ln k}
  {\int \Delta^2(k) W^2(kR)\,\rmd\ln k}.
\end{equation}
For the quasars, the minimum mass (which for a steep mass function can be taken as a typical mass) is $M\sim (2-6)\times 10^{12}h^{-1}M_\odot$ 
\cite{2007AJ....133.2222S}, which gives $k_{\rm eff}=(1.3-1.8)h\,$Mpc$^{-1}$.  This range spans only 0.3 e-folds and in what follows we will thus take 
the central value, $k_{\rm eff}=1.5h\,$Mpc$^{-1}$.

One can make a similar estimate for the Eriksen \etal \cite{2007ApJ...660L..81E}; we find in \ref{sec:cmbscale} that $k_{\rm 
eff}=0.0033h\,$Mpc$^{-1}$.  The Eriksen \etal constraint and our constraint are thus separated by 6.1 e-folds.  This strongly constrains inflationary explanations for 
the power asymmetry: the asymmetry amplitude $A$ must decrease by a factor of at least a few ($\sim 7$\% to $<\sim1$\%) during only 6.1 e-folds of expansion.  This 
contrasts with the usual expectation from inflationary models of nearly scale-invariant spectra.  One could imagine features in the inflationary potential that 
strongly break scale invariance; however there is no motivation to place such features in precisely this range of $\ln k$, and moreover such a model would have to 
avoid ``breaking'' the successes of standard inflation in predicting the CMB acoustic peak regime and large-scale galaxy clustering.  A detailed study of such 
constraints is presented in a companion paper \cite{EHK09}.

\section{Discussion}

One of the intriguing products of the WMAP mission was the detection of large-scale anomalies in the CMB.  If these anomalies are ``real'' (as
opposed to statistical fluctuations or systematics) then their implications are revolutionary.  The anomalies are thus in need of confirmation,
both by future CMB data that could detect them at higher signal-to-noise ratio (e.g. {\it Planck}) and just as importantly by independent probes
of the same underlying physics.  In particular, the dipolar asymmetry in the CMB power spectrum reported by Eriksen \etal\ \cite{2007ApJ...660L..81E},
if interpreted as a spatial variation of the amplitude of primordial fluctuations, makes predictions not just for the CMB but also for
large-scale structure: if there is truly a large-scale gradient $\nabla\sigma_8$ across the Universe, then there shuld be a corresponding gradient
in the number density of highly biased objects.  We have searched for such a gradient using the SDSS $z\ge 2.9$ quasars and found a null result.
We find that the dipole ${\bi p}=r_{\rm lss}\nabla\ln\sigma_8$ is no more than 0.027 (99\% posterior probability), i.e. any smooth gradient in the
amplitude of fluctuations is no more than 2.7 percent per present-day horizon radius.  Tighter results are obtained if one forces the gradient to
be in the same direction as that reported by Eriksen \etal\ \cite{2007ApJ...660L..81E}.

This result rules out the simple curvaton-gradient model \cite{2008PhRvD..78l3520E} in which the power asymmetry in the CMB comes from a spatial gradient
in the curvaton field of a two-field inflation model.  Such a model requires $p\approx 0.11$ to reproduce the reported $\pm11$\% variation in the CMB amplitude
across the sky \cite{2007ApJ...660L..81E}.  This does not necessarily contradict the Eriksen \etal\ \cite{2007ApJ...660L..81E} result: there may be
more complicated models with a scale-dependent gradient in small-scale power, i.e. where the spatial gradient of the local $\ln P(k)$ depends on $k$.
Such models, including some that predict much smaller dipoles in the quasar distribution and the degree-scale CMB fluctuation amplitudes,
are explored in the companion paper by Erickcek \etal \cite{EHK09}.

\ack

CH wishes to thank Y Shen for providing his quasar sample and associated mask files.
CH also wishes to thank A Erickcek, M Kamionkowski, and Y Shen for their useful comments on the paper.

CH is supported by the US Department of Energy under contract DE-FG03-02-ER40701, the National Science Foundation under contract AST-0807337, and the 
Alfred P Sloan Foundation.

Funding for the Sloan Digital Sky Survey (SDSS) and SDSS-II has been provided by the Alfred P Sloan Foundation, the Participating Institutions, 
the National Science Foundation, the US Department of Energy, the National Aeronautics and Space Administration, the Japanese Monbukagakusho, and the 
Max Planck Society, and the Higher Education Funding Council for England. The SDSS Web site is http://www.sdss.org/.

The SDSS is managed by the Astrophysical Research Consortium (ARC) for the Participating Institutions. The Participating Institutions are the 
American Museum of Natural History, Astrophysical Institute Potsdam, University of Basel, University of Cambridge, Case Western Reserve University, The 
University of Chicago, Drexel University, Fermilab, the Institute for Advanced Study, the Japan Participation Group, The Johns Hopkins University, the 
Joint Institute for Nuclear Astrophysics, the Kavli Institute for Particle Astrophysics and Cosmology, the Korean Scientist Group, the Chinese Academy 
of Sciences (LAMOST), Los Alamos National Laboratory, the Max-Planck-Institute for Astronomy (MPIA), the Max-Planck-Institute for Astrophysics (MPA), 
New Mexico State University, Ohio State University, University of Pittsburgh, University of Portsmouth, Princeton University, the United States Naval 
Observatory, and the University of Washington.

\appendix

\section{Effective scale of CMB power asymmetry}
\label{sec:cmbscale}

Eriksen \etal \cite{2007ApJ...660L..81E} measured a dipole asymmetry in the CMB power spectrum in the range of multipoles $2\le\ell\le \ell_{\rm 
max}\sim 40$.  The upper limit $\ell_{\rm max}$ was set by the requirement to be oversampled on the {\sc Healpix} resolution 4 ($N_{\rm side}=16$) 
pixelization scheme \cite{2005ApJ...622..759G}.  Our goal here is to estimate the effective wavenumber of the Eriksen \etal measurement.

In the Sachs-Wolfe regime, and neglecting the small contribution from the integrated Sachs-Wolfe effect, the temperature perturbation is given by $\Theta(\hat\bi n) = 
- \zeta(r_{\rm lss}\hat\bi n)/5$ where $\zeta$ is the primordial curvature 
perturbation.  Therefore the CMB power spectrum at multipole $\ell$ is obtained by
\begin{equation}
C_\ell = \frac{4\pi}{25}\int j_\ell^2(kr_{\rm lss}) \Delta^2_\zeta(k) \rmd\ln k,
\end{equation}
where $\Delta^2_\zeta(k)$ is the primordial curvature power spectrum.  [See, e.g. Eqs.~(5.27) and (5.39) of 
Ref.~\cite{2000cils.book.....L}.]  For smooth power spectra and $\ell\gg 
1$, we can use the Wentzel-Kramers-Brillouin approximation for the spherical Bessel function,
\begin{equation}
j_\ell(x) \rightarrow x^{-1/2}(x^2-\ell^2)^{-1/4}\cos\varphi(x)
\end{equation}
for $x>\ell$, where $\varphi(x)$ is a rapidly oscillating phase.  (For $x<\ell$ the spherical Bessel function goes rapidly to zero.)  With this 
approximation, and taking $\langle\cos^2\varphi(x)\rangle\rightarrow\frac12$, $C_\ell$ simplifies to
\begin{equation}
C_\ell \approx \frac{2\pi}{25\ell^2}\int y^{-1}(y^2-1)^{-1/2} \Delta^2_\zeta(k) \rmd\ln k,
\end{equation}
where $y=x/\ell=kr_{\rm lss}/\ell$.
The fractional dipole asymmetry $\delta\ln C_\ell$ of $C_\ell$ is given by:
\begin{equation}
\delta\ln C_\ell \approx
\frac{
\int y^{-1}(y^2-1)^{-1/2} \Delta^2_\zeta(k)\delta\ln \Delta^2_\zeta(k) \rmd\ln k
}{
\int y^{-1}(y^2-1)^{-1/2} \Delta^2_\zeta(k) \rmd\ln k
}.
\end{equation}
This equation can be simplified for the case of a near scale-invariant reference spectrum, $\Delta^2_\zeta(k)=$constant:
\begin{equation}
\delta\ln C_\ell \approx
\frac{
\int_{y=1}^\infty y^{-1}(y^2-1)^{-1/2} \delta\ln \Delta^2_\zeta(k) \rmd\ln y
}{
\int_{y=1}^\infty y^{-1}(y^2-1)^{-1/2} \rmd\ln y
}.
\end{equation}
The integral in the denominator evaluates to 1 (with the substitution $y=\sec\beta$).

In order to make further progress, we assume that $\delta\ln 
\Delta^2_\zeta(k)$ varies 
smoothly with $\ln k$, as 
typical during inflation models, i.e. we assume:
\begin{equation}
\delta\ln \Delta^2_\zeta(k) = B_1 + B_2\ln (kr_{\rm lss}) = B_1+B_2\ln\ell + B_2\ln y.
\end{equation}
Only the $\ln y$ part of the $C_\ell$ integral is nontrivial,
\begin{eqnarray}
\delta\ln C_\ell &\approx&
B_1 + B_2\ln\ell + B_2
\int_{y=1}^\infty y^{-1}(y^2-1)^{-1/2}\ln y\,\rmd\ln y
\nonumber \\
&=& B_1 + B_2\ln\ell + B_2(1-\ln2).
\end{eqnarray}
\{The integral over $y$ can be solved by the substitution $y=(1-\rho^2)^{-1/2}$, which turns it into 
$-\frac12\int_0^1[\ln(1-\rho)+\ln(1+\rho)]\rmd\rho$.  The ln functions are then integrated by parts.\}

The Eriksen \etal \cite{2007ApJ...660L..81E} dipole asymmetry measurement fits a single amplitude for the power asymmetry across their entire range of 
$\ell$.  In the limit where all multipoles up to $\ell_{\rm max}$ are signal-dominated and those above $\ell_{\rm max}$ are noise-dominated, the 
Eriksen \etal asymmetry parameter $A$ will receive equal weight from all modes, i.e.
\begin{equation}
A = \frac12 \langle \delta\ln C_\ell \rangle_{\rm modes}
 \approx \frac12 \left[ B_1 + B_2\langle\ln\ell\rangle_{\rm modes} + B_2(1-\ln2) \right].
\end{equation}
(The factor of 1/2 comes from the fact that Eriksen \etal measured an amplitude asymmetry rather than a power asymmetry.)  Since the number of modes 
in any region of the sky and any $\ell$ range is proportional to $\ell\,\rmd\ell$, we have, for $\ell_{\rm max}\gg 1$,
\begin{equation}
\langle\ln\ell\rangle_{\rm modes} = \frac{\int_0^{\ell_{\rm max}} \ell\ln\ell\,\rmd\ell}{\int_0^{\ell_{\rm max}} \ell\,\rmd\ell}
= \ln \ell_{\rm max}-\frac12.
\end{equation}
(The integral in the numerator is evaluated using repeated integration by parts.)  Thus:
\begin{equation}
A = \frac12 \langle \delta\ln C_\ell \rangle_{\rm modes}
 \approx \frac12 \left[ B_1 + B_2\left(\ln\ell_{\rm max}+\frac12-\ln2\right) \right].
\end{equation}
This is the same as the amplitude asymmetry $\frac12\delta\ln\Delta^2_\zeta(k_{\rm eff})$ at the effective scale:
\begin{equation}
k_{\rm eff} = \frac{\rme^{1/2}\ell_{\rm max}}{2r_{\rm lss}},
\end{equation}
where $\rme=2.718...$ is the base of the natural logarithm.

For the specific case of Eriksen \etal, $r_{\rm lss}=1.0\times 10^4h^{-1}\,$Mpc and $\ell_{\rm max}=40$, so $k_{\rm eff}=0.0033h\,$Mpc$^{-1}$.
Similarly, for Hoftuft \etal \cite{2009ApJ...699..985H},
$k_{\rm eff}=0.0053h\,$Mpc$^{-1}$ ($\ell_{\rm max}=64$) or
$k_{\rm eff}=0.0066h\,$Mpc$^{-1}$ ($\ell_{\rm max}=80$).

\section*{References}

\bibliography{gradsig}

\end{document}